\def\ARXIVVERSION{1}

\ifdefined\PAPERCLASSLOADED
\else
  \ifdefined\ARXIVVERSION
    \documentclass[journal]{IEEEtran}
  \else
    \documentclass[manuscript,review,anonymous,screen]{acmart}
  \fi
\fi

\AtBeginDocument{%
  \providecommand\BibTeX{{\normalfont B\kern-0.5em{\scshape i\kern-0.25em b}\kern-0.8em\TeX}}}

\ifdefined\ARXIVVERSION
  \usepackage{amsthm}
  \usepackage[table]{xcolor}
  \usepackage{graphicx}
  \usepackage{hyperref}
  \hypersetup{hidelinks}
  \usepackage[caption=false,font=footnotesize]{subfig}
  \usepackage{placeins}
  \providecommand{\Description}[1]{}
\else
  \setcopyright{acmlicensed}
  \acmJournal{TODAES}
  \acmYear{2026}
  \acmVolume{0}
  \acmNumber{0}
  \acmArticle{0}
  \acmMonth{1}
  \acmDOI{10.1145/nnnnnnn.nnnnnnn}
  \usepackage{subcaption}
\fi

\usepackage{amsmath,amsfonts}
\DeclareMathAlphabet\mathbfcal{OMS}{cmsy}{b}{n}

\newcommand{\mat}[1]{\mathbf{#1}}
\usepackage{algorithm}
\usepackage{algpseudocode}
\usepackage{array}
\usepackage{booktabs}
\usepackage{multirow}
\usepackage{colortbl}
\usepackage{float}

\definecolor{tabBandYield}{HTML}{D9E8F5}
\definecolor{tabBandP}{HTML}{DEEBD9}
\definecolor{tabRowAlt}{HTML}{F4F4F4}
\definecolor{deltapos}{HTML}{1B7837}
\definecolor{deltaneg}{HTML}{C0392B}

\newtheorem{theorem}{Theorem}
\newtheorem{proposition}{Proposition}
\theoremstyle{definition}
\newtheorem{assumption}{Assumption}

\ifdefined\ARXIVVERSION
  \newenvironment{paperwidetable}[1][t]{\begin{table*}[#1]}{\end{table*}}
  \newenvironment{paperwidefigure}[1][t]{\begin{figure*}[#1]}{\end{figure*}}
  \newenvironment{paperappendixtable}{\begin{table}[t]}{\end{table}}
\else
  \newenvironment{paperwidetable}[1][t]{\begin{table}[#1]}{\end{table}}
  \newenvironment{paperwidefigure}[1][t]{\begin{figure}[#1]}{\end{figure}}
  \newenvironment{paperappendixtable}{\begin{table}[H]}{\end{table}}
\fi

\begin{document}

\ifdefined\ARXIVVERSION
\title{Simulation-Efficient Analog Circuit Yield Optimization via
       Monte Carlo Zeroth-Order Gradient Estimation}

\author{Liyan~Tan, Yequan~Zhao, Ben~F.~Jamroz, Ari~Feldman, and Zheng~Zhang%
\thanks{L. Tan, Y. Zhao, and Z. Zhang are with the Department of
Electrical and Computer Engineering, University of California, Santa
Barbara, CA, USA (e-mail: liyan\_tan@ucsb.edu; yequan\_zhao@ucsb.edu;
zhengzhang@ece.ucsb.edu).}%
\thanks{B. F. Jamroz and A. Feldman are with the National Institute of
Standards and Technology, Boulder, CO, USA (e-mail:
benjamin.jamroz@nist.gov; ari.feldman@nist.gov).}}

\markboth{Preprint}{Tan \MakeLowercase{\textit{et al.}}: Simulation-Efficient Analog Circuit Yield Optimization}
\maketitle

\begin{abstract}
Yield optimization under process variation is expensive because each candidate design must be evaluated across many Monte Carlo SPICE\footnote{Certain commercial software is identified in this paper to foster understanding. Such identification does not imply recommendation or endorsement by the National Institute of Standards and Technology, nor does it imply that the software identified is necessarily the best available for the purpose.} samples. The resulting finite-sample yield is also piecewise constant in the design parameters, providing little local information for optimization. We introduce zeroth-order Monte Carlo stochastic gradient descent (ZO-MC-SGD), a black-box method that converts continuous specification margins into stochastic descent directions. Each update evaluates opposite design perturbations under shared process samples, allowing a small simulation batch to estimate a local direction without differentiating SPICE or fitting a global surrogate model. A Spearman rank-correlation test checks that the margin-based loss orders designs consistently with empirical yield. We prove that the estimator is unbiased for a Gaussian-smoothed surrogate and derive variance and sample-complexity bounds with no explicit dependence on process dimension. Across five analog circuit benchmarks with up to 30 design variables and 42 process variables, ZO-MC-SGD reaches a mean yield of 0.95 on four circuits within 50--200 simulations and the empirical yield ceiling on the fifth. Relative to the best of five black-box and learning-based baselines, it reduces the required simulation budget by up to a factor of eight.
\end{abstract}

\else
\title{Sample-Efficient Yield Optimization of Analog Circuits via
       Stochastic Zeroth-Order Methods}

\author{Liyan Tan}
\orcid{0009-0006-3990-1203}
\affiliation{%
  \institution{University of California, Santa Barbara}
  \department{Department of Electrical and Computer Engineering}
  \city{Santa Barbara}
  \state{CA}
  \country{USA}}
\email{liyan\_tan@ucsb.edu}

\author{Yequan Zhao}
\orcid{0009-0004-2785-6789}
\affiliation{%
  \institution{University of California, Santa Barbara}
  \department{Department of Electrical and Computer Engineering}
  \city{Santa Barbara}
  \state{CA}
  \country{USA}}
\email{yequan\_zhao@ucsb.edu}

\author{Ben F. Jamroz}
\orcid{0000-0002-5498-1137}
\affiliation{%
  \institution{National Institute of Standards and Technology}
  \city{Boulder}
  \state{CO}
  \country{USA}}
\email{benjamin.jamroz@nist.gov}

\author{Ari Feldman}
\orcid{0000-0002-9603-4031}
\affiliation{%
  \institution{National Institute of Standards and Technology}
  \city{Boulder}
  \state{CO}
  \country{USA}}
\email{ari.feldman@nist.gov}

\author{Zheng Zhang}
\orcid{0000-0002-2292-0030}
\affiliation{%
  \institution{University of California, Santa Barbara}
  \department{Department of Electrical and Computer Engineering}
  \city{Santa Barbara}
  \state{CA}
  \country{USA}}
\email{zhengzhang@ece.ucsb.edu}

\renewcommand{\shortauthors}{Tan et al.}

\begin{abstract}
Process variation makes yield a central concern in analog circuit design because a circuit must satisfy all specifications across manufacturing variations, not only at the nominal operating point. Direct yield optimization is difficult, however, because finite-sample Monte Carlo yield estimates are non-smooth with respect to design parameters, while accurate estimates can require many costly SPICE simulations. To address these challenges, zeroth-order Monte Carlo stochastic gradient descent (ZO-MC-SGD) is introduced as a black-box method for sample-efficient yield optimization. The method replaces the binary pass/fail objective with a smooth surrogate based on specification margins and estimates descent directions from a small number of perturbed SPICE simulations. A rank-correlation criterion checks whether the surrogate preserves the design ordering induced by empirical yield. The resulting gradient estimator is analyzed theoretically, with guarantees on its accuracy and sample complexity. Across five analog circuit benchmarks, ZO-MC-SGD reaches a mean yield of 0.95 on four circuits within 50--200 SPICE simulations and the empirical yield ceiling on the fifth. It reduces the required simulation budget by up to a factor of eight relative to the best baseline; several of the five black-box and learning-based baselines fail to reach the target even at substantially larger budgets.
\end{abstract}



\maketitle
\fi

\section{Introduction}
\label{sec:intro}
Process variations are unavoidable in nanometer-scale CMOS
fabrication. Random dopant fluctuations, line-edge roughness, and
oxide-thickness variations perturb device characteristics and
propagate to circuit-level performance~\cite{pelgrom1989,
drennan2003understanding}. A circuit that meets every specification
at the nominal process point may therefore violate its gain,
bandwidth, or power requirement after fabrication. Its \emph{yield},
defined as the probability that a fabricated instance satisfies all
specifications under process variation, measures whether nominal
performance survives these variations. Yield rather than nominal
performance is consequently the relevant objective for a
manufacturable design.

However, making yield the design objective creates two computational
difficulties. First, an
empirical yield estimate averages binary pass/fail outcomes. For a
fixed set of Monte Carlo (MC) process samples, a small change in the
design often leaves every outcome unchanged, making the estimate
piecewise constant and unable to provide a useful descent direction.
Second, distinguishing a high-yield design from a nearby candidate
requires many SPICE simulations, and an optimizer repeats this
estimation across many designs. The cost becomes especially severe
in distribution tails: rare-event analyses such as SRAM failure
estimation can require millions of Monte Carlo
samples~\cite{singhee2009statistical,kanj2006mixture}. Direct yield
optimization must thus
extract useful search information without repeatedly paying for a
high-accuracy yield estimate.

Existing approaches address one side of this problem but leave the
other. Adjoint methods provide gradients efficiently, but require
derivatives from within the simulator that are often unavailable in
proprietary flows. Black-box methods avoid this access requirement.
Bayesian optimization and its trust-region variants learn a model
from evaluated designs~\cite{snoek2012practical,frazier2018tutorial,
wang2017efficient,lyu2018efficient,lyu2018batch,
eriksson2019turbo}, while evolutionary strategies and particle
swarm evaluate an initial population~\cite{hansen2016cma,
kennedy1995particle}. These initial designs consume a substantial
part of a budget of only a few hundred high-fidelity simulations.
Reinforcement-learning sizing methods require per-circuit policy
training over many rollouts~\cite{settaluri2020autockt,
shi2022robustanalog}; moreover, their rewards typically measure
robustness across selected process corners rather than Monte Carlo
yield. Together, these limitations leave a practical question: can
an optimizer obtain a useful design direction without differentiating
SPICE or repeatedly estimating yield to high accuracy?

Zeroth-order Monte Carlo stochastic gradient descent (ZO-MC-SGD) takes
this route. Zeroth-order optimization retains the
update structure of gradient descent, but estimates the gradient from
finite differences of forward simulator evaluations rather than from
analytic or adjoint derivatives~\cite{nesterov2017random,duchi2015optimal,
berahas2022theoretical}. At each iteration, ZO-MC-SGD perturbs the
current design along random directions and evaluates the paired
designs over a small batch of process samples. It therefore obtains a
stochastic descent direction using only black-box SPICE simulations.

The objective used in these finite differences is equally important.
Applying them directly to empirical yield would recover the same
piecewise-constant signal described above. ZO-MC-SGD instead
constructs a smooth loss surrogate from specification margins and
uses rank correlation to check that lower loss corresponds to higher
yield. The resulting estimator is unbiased for the gradient of the
smoothed surrogate, and its simulation complexity has no explicit
dependence on the process dimension. We establish convergence to
stationarity of the surrogate and evaluate the returned designs by an
independent Monte Carlo yield estimate.

The paper presents the following contributions:
\begin{itemize}
\item \textbf{Algorithm.} We develop ZO-MC-SGD, a stochastic
  zeroth-order method that optimizes a yield-aligned, smooth
  specification-margin surrogate using paired Monte Carlo gradient
  estimates.
\item \textbf{Theoretical analysis.} We prove that the gradient
  estimator is unbiased for the smoothed surrogate and derive bounds
  with no explicit process-dimension dependence. Synthetic tests
  verify the estimator against analytic ground truth.
\item \textbf{Numerical results.} On five analog integrated-circuit design benchmarks
  with up to 42 process variables, ZO-MC-SGD reaches a mean yield of
  0.95 on four circuits within 50--200 simulations and the empirical
  yield ceiling on the fifth, reducing the required simulation budget
  by up to a factor of eight relative to the best baseline. Code is
  available at \url{https://anonymous.4open.science/r/ZO-MC-SGD}.
\end{itemize}

\section{Background and Related Work}
\label{sec:setting}

Yield optimization involves two distinct tasks: evaluating circuit
performance under process variation and deciding how to update the
design from costly simulator evaluations. Monte Carlo simulation
addresses the first task by estimating yield at a candidate design, but
it does not prescribe how the next design should be chosen. Existing
optimization methods differ mainly in how they obtain this design
update. Model-based methods construct an explicit representation of
performance or yield and optimize it to guide the search. Learning-based
methods instead train a sizing policy from repeated simulator
interactions. Zeroth-order methods estimate a local descent direction
directly from perturbed function evaluations. These different uses of
simulation data determine both how the search proceeds and how much
simulation cost is required to obtain useful design updates.

\subsection{Monte Carlo Yield Estimation}
\label{sec:setting:yield}

Yield measures how reliably a circuit meets its specifications under
process variation. For a design
$\mat{x}\in\mathcal{X}\subset\mathbb{R}^n$, let
$\boldsymbol{\xi}$ denote the process variables with density
$\rho(\boldsymbol{\xi})$, and let
$\mat{f}(\mat{x},\boldsymbol{\xi})$ collect the resulting performance
metrics. If $\mathcal{S}$ denotes the feasible specification region, the
yield is the probability that all specifications are satisfied,
\begin{equation}
Y(\mat{x})
\;=\;
\mathbb{P}_{\boldsymbol{\xi}\sim\rho}
\!\left[
\mat{f}(\mat{x},\boldsymbol{\xi})\in\mathcal{S}
\right]
\;=\;
\int
\mathbf{1}\!\left[
\mat{f}(\mat{x},\boldsymbol{\xi})\in\mathcal{S}
\right]
\rho(\boldsymbol{\xi})\,
\mathrm{d}\boldsymbol{\xi},
\label{eq:rw:yield}
\end{equation}
where $\mathbf{1}(\cdot)$ is the pass/fail indicator. Yield optimization
therefore seeks a design that maximizes $Y(\mat{x})$ over
$\mathcal{X}$.

This probability is rarely available in closed form. In practice, it is
estimated from $N$ SPICE simulations using independent and identically
distributed (i.i.d.) process samples,
\begin{equation}
\hat{Y}(\mat{x})
\;=\;
\frac{1}{N}
\sum_{j=1}^{N}
\mathbf{1}\!\left[
\mat{f}(\mat{x},\boldsymbol{\xi}^{(j)})\in\mathcal{S}
\right],
\qquad
\boldsymbol{\xi}^{(j)}
\overset{\mathrm{i.i.d.}}{\sim}\rho.
\label{eq:rw:mc}
\end{equation}
Monte Carlo thus makes yield directly evaluable at any candidate
design, but it provides little local information for updating that
design. With a fixed sample set, $\hat{Y}(\mat{x})$ changes only when a
sample crosses a specification boundary, so the estimate is piecewise
constant over large regions of the design space. Increasing $N$ reduces
sampling noise, but the estimation error decreases only as $N^{-1/2}$
and the simulation cost becomes especially high when failures are rare.

Variance-reduction methods reduce the cost of obtaining a reliable yield
estimate. Antithetic variates correlate samples to reduce estimator
variance~\cite{hammersley1956new}, while importance sampling and
statistical blockade concentrate simulations near failure
events~\cite{singhee2009statistical,kanj2006mixture}. These methods
improve yield estimation at a fixed candidate design, but they do not
determine how the resulting evaluations should be used to update the
design. That question is handled by the optimization methods discussed
next.

\subsection{Model-Based Yield Optimization}
\label{sec:setting:surrogate}

Model-based methods address the search problem by constructing a
tractable representation of circuit performance or yield from simulator
evaluations and using that representation to guide the design search.
Bayesian optimization maintains a probabilistic model of the objective
and balances promising designs against uncertain regions
\cite{snoek2012practical,frazier2018tutorial}. Weighted and batch
formulations adapt this strategy to analog circuit sizing
\cite{lyu2018efficient,lyu2018batch}, while truncated subspaces and
local trust regions extend it to higher-dimensional design spaces
\cite{wang2024tssbo,eriksson2019turbo}.

For variation-aware design, the model can capture the effect of process
variation rather than nominal performance alone. Bayesian yield
optimization and freeze--thaw schemes model yield directly and allocate
simulation effort across candidate designs
\cite{wang2017efficient,wang2022freezethaw}. Gaussian-process
classification instead learns the pass/fail boundary
\cite{wang2022gpc}. Bayesian neural networks model yields and circuit
performances jointly across multiple process, voltage, and temperature
(PVT) corners and use the learned
model within a Bayesian optimization framework
\cite{guo2023bnn}. Contextual robust optimization jointly models design
and perturbation variables with Gaussian processes and searches for
solutions that remain robust under prescribed variations
\cite{huang2025croza}. Polynomial-chaos methods take a different route:
they construct an explicit approximation of stochastic circuit
performance over the process variables and use it to solve a
chance-constrained design problem
\cite{cui2020chance,he2021pobo}.

These approaches turn simulator evaluations into design guidance through
an explicit model. Their simulation cost depends on how much data are
needed to construct an informative representation, with additional
computation spent updating the model and solving the associated
acquisition, robust, or constrained optimization problem.

\subsection{Variation-Aware Learning-Based Design}
\label{sec:setting:rl}

Learning-based methods provide another route by using simulator
interactions to train a policy that maps the current circuit state to
sizing actions. AutoCkt established this sequential-decision formulation
for nominal analog sizing, with rewards based on specification
satisfaction and constraint violation
\cite{settaluri2020autockt}. Through repeated simulator interactions,
the policy learns which sizing actions improve the design.

Recent methods extend this policy-based formulation to variation-aware
design. RobustAnalog shares experience across corner tasks and prunes
redundant ones to reduce simulation cost
\cite{shi2022robustanalog}. PVTSizing combines multi-task reinforcement
learning with trust-region Bayesian optimization
\cite{kong2024pvtsizing}, while RoSE-Opt incorporates domain knowledge
and Bayesian-optimization warm starts
\cite{cao2024roseopt}. GLOVA instead introduces a risk-sensitive
objective that accounts for global and local variation
\cite{kim2025glova}. These methods learn sizing policies that improve
specification satisfaction or robustness under variation rather than
optimizing Monte Carlo yield directly. Training typically requires many
simulator interactions and is repeated for each circuit or topology.

\subsection{Zeroth-Order Optimization}
\label{sec:setting:zo}

Zeroth-order (ZO) optimization enables gradient-based search using
function evaluations alone. Instead of differentiating the objective,
a ZO method perturbs the current iterate, observes the resulting
function values, and uses their differences to estimate a local descent
direction. Randomized finite-difference estimators and stochastic
approximation methods such as simultaneous perturbation stochastic
approximation (SPSA) established this idea in
derivative-free optimization~\cite{spall1992multivariate,
flaxman2005bandit,nesterov2017random,duchi2015optimal,
ghadimi2013stochastic,berahas2022theoretical}. Unlike population-based
methods such as evolution strategies and particle swarm optimization,
ZO explicitly constructs a gradient estimate that can be used in a
gradient-based update~\cite{hansen2016cma,kennedy1995particle}.

The same idea has been increasingly used in machine learning when
gradients are unavailable, hidden behind a black-box interface, or
expensive to obtain~\cite{liu2020primer}. More recently, ZO methods have
enabled memory-efficient language-model fine-tuning, where forward
evaluations replace the backward pass needed for exact
gradients~\cite{malladi2023fine,
tan2026grzogrouprelativezerothorderoptimization}. These applications
highlight the main appeal of ZO: retaining gradient-based updates while
requiring only function values.

Circuit simulation presents a closely related black-box setting because
commercial simulators provide performance evaluations but generally do
not expose gradients. ZOAF applies two-point zeroth-order estimates to
analog/RF circuit sizing and obtains descent directions from a small
number of simulator evaluations
\cite{tan2026zoafefficientzerothorderoptimization}. Its optimization
setting is nominal and deterministic: each design is evaluated at a fixed
process condition using a deterministic figure of merit.

Yield optimization introduces a different objective. The quantity of
interest is a probability over random process realizations, while its
finite-sample Monte Carlo estimate is piecewise constant with respect to
the design variables. Directly applying a zeroth-order update to this
empirical yield therefore provides little useful search information. The
method developed next addresses this difficulty by combining a smooth
loss derived from specification margins with Monte Carlo zeroth-order
gradient estimation, allowing local descent directions to be obtained
from black-box simulations under process variation.

\section{Method}
\label{sec:method}

Zeroth-order optimization can recover a local search direction from
black-box function evaluations, but empirical yield is a poor quantity
to perturb. Each Monte Carlo sample is reduced to a binary pass/fail
outcome, so designs with very different performance margins can produce
the same yield estimate. With a fixed sample set, that estimate is also
piecewise constant over large regions of the design space.

ZO-MC-SGD addresses this difficulty by changing the signal used for
optimization while keeping yield as the final design objective. Each
SPICE simulation is first converted into continuous specification
margins, which are aggregated into a smooth loss whose ordering can be checked
against empirical yield. Paired perturbations of this loss are then averaged
over process samples to obtain a stochastic zeroth-order descent
direction. The method therefore separates two questions: what
information should be retained from each simulation, and how should
that information be turned into a design update.

Using the notation of Section~\ref{sec:setting:yield}, let
$\ell(\mat{x},\boldsymbol{\xi})$ denote the per-sample loss constructed
below. ZO-MC-SGD optimizes its expectation,
\begin{equation}
\min_{\mat{x}\in\mathcal{X}} F(\mat{x}),
\qquad
F(\mat{x})
:=
\mathbb{E}_{\boldsymbol{\xi}\sim\rho}
\big[\ell(\mat{x},\boldsymbol{\xi})\big].
\label{eq:problem}
\end{equation}
The yield $Y(\mat{x})$ in~\eqref{eq:rw:yield} remains the quantity used
to evaluate the returned design; $F$ supplies the optimization signal.
The next section constructs this signal from specification margins,
followed by the Monte Carlo zeroth-order update used to descend it.

\subsection{From Specification Margins to a Yield-Aligned Loss}
\label{sec:method:loss}

A SPICE simulation contains considerably more information than the
yield indicator retains. The indicator treats a sample that barely
misses a specification in the same way as one that misses it by a wide
margin, and similarly discards the amount of headroom once a
specification is satisfied. These distances from the specification
boundaries provide useful local information for optimization. The loss
therefore begins with a signed margin for each circuit metric.

Because different metrics have different units and numerical scales,
their raw margins cannot be combined directly. Gain is expressed in
decibels, unity-gain bandwidth in decades of hertz, and static power in
milliwatts before the margins are formed. Let
$f_k(\mat{x},\boldsymbol{\xi})$ and $\tau_k$ denote the resulting scaled
metric and its specification threshold. The signed shortfall is
\begin{equation}
\delta_k(\mat{x}, \boldsymbol{\xi}) =
\begin{cases}
\tau_k - f_k(\mat{x}, \boldsymbol{\xi}),
& \text{spec is}\ f_k \geq \tau_k,\\[2pt]
f_k(\mat{x}, \boldsymbol{\xi}) - \tau_k,
& \text{spec is}\ f_k \leq \tau_k.
\end{cases}
\end{equation}
Thus, $\delta_k>0$ measures a specification violation, whereas
$\delta_k<0$ measures the available margin beyond the boundary.

A useful optimization signal should penalize violations without
discarding information near the specification boundary. We therefore
map each signed shortfall through a softplus,
\begin{equation}
\sigma_k(\mat{x}, \boldsymbol{\xi}; \alpha_k)
=
\frac{1}{\alpha_k}
\log\!\big(
1 + e^{\alpha_k\delta_k(\mat{x},\boldsymbol{\xi})}
\big).
\label{eq:smoothed-spec}
\end{equation}
The penalty grows nearly linearly on the violated side and decays
smoothly as margin is gained. The sharpness $\alpha_k$ controls how far
a specification continues to influence the loss around its boundary:
larger values approach a hinge, while smaller values continue to reward
margin farther into the feasible region.

The per-sample loss combines these specification penalties,
\begin{equation}
\ell(\mat{x}, \boldsymbol{\xi})
=
\sum_{k=1}^{K_{\mathrm{spec}}}
w_k\,\sigma_k(\mat{x},\boldsymbol{\xi};\alpha_k)
-
\gamma f_{\mathrm{gain}}(\mat{x},\boldsymbol{\xi}).
\label{eq:loss}
\end{equation}
The weights $w_k$ balance the scaled specifications, and the final term
provides a mild preference for gain headroom rather than stopping
exactly at the gain threshold.

\label{sec:method:alpha}
Smoothness alone, however, does not make the loss a useful surrogate
for yield. The loss is the optimization target, but yield is the score;
their numerical values need not agree, but lower loss should generally
identify higher-yield designs. We assess this relationship using the
Spearman rank correlation $\rho_s$ between mean loss and empirical
yield over a fixed calibration set of $N_{\mathrm{cal}}$ perturbed
designs. When circuit-specific calibration is used, candidate
sharpness settings are evaluated on the same set and the setting with
the most negative $\rho_s$ is selected. A loss is considered
sufficiently aligned when $\rho_s\leq-0.7$.

The rank test evaluates the ordering induced by the complete loss after
scaling, weighting, and aggregation. Section~\ref{sec:exp:objalign}
examines the sensitivity to the sharpness choice and the use of a
shared default.

\subsection{Monte Carlo Zeroth-Order Descent}
\label{sec:method:estimators}

Once the yield-aligned loss is fixed, the remaining task is to turn its
black-box evaluations into a design update under process variation. At
iteration $t$, ZO-MC-SGD draws $K$ process samples
$\{\boldsymbol{\xi}^{(i)}\}_{i=1}^{K}$ from $\rho$ and $K$ random
directions $\mat{v}^{(i)}\sim\mathcal{N}(0,I_n)$. For each pair, the
design is perturbed in opposite directions while the process
realization is held fixed, and the resulting differences are averaged:
\begin{equation}
\hat{\mat{g}}^{\mathrm{MC}}_t
=
\frac{1}{K}
\sum_{i=1}^{K}
\frac{
\ell(\mat{x}_t+\epsilon\mat{v}^{(i)},\boldsymbol{\xi}^{(i)})
-
\ell(\mat{x}_t-\epsilon\mat{v}^{(i)},\boldsymbol{\xi}^{(i)})
}{2\epsilon}
\,\mat{v}^{(i)}.
\label{eq:opt1}
\end{equation}
The two perturbations in each pair share the same process sample, so
their difference isolates the effect of the design perturbation
\cite{lecuyer1994efficiency,glasserman2003monte}. Averaging $K$ such
pairs gives one stochastic gradient estimate at a cost of $2K$ SPICE
simulations, with $\epsilon$ setting the perturbation scale and $K$
controlling the amount of Monte Carlo averaging.

The estimate in~\eqref{eq:opt1} can then be used in a standard
stochastic-gradient update. We first map each physical design box
affinely onto $\mathcal{X}=[0,1]^n$, giving the perturbation radius
$\epsilon$ and optimization step a common coordinate scale, and project
each update back onto the box:
\begin{equation}
\mat{x}_{t+1}
=
\Pi_{\mathcal{X}}
\!\left(
\mat{x}_t-\eta\hat{\mat{g}}^{\mathrm{MC}}_t
\right).
\label{eq:descent}
\end{equation}
In practice, circuit parameters can have very different sensitivities,
so ZO-MC-SGD uses Adam to rescale the update across coordinates
\cite{kingma2014adam}. Algorithm~\ref{alg:vanilla-zo} combines this
descent with a short random-search warm start. The warm start is counted
toward the SPICE optimization budget $B$; after it, each gradient estimate costs
$2K$ simulations, and the remaining budget determines how many descent
steps can be taken.

\begin{algorithm}[t]
\caption{ZO-MC-SGD}
\label{alg:vanilla-zo}
\begin{algorithmic}[1]
\Require loss $\ell$, process density $\rho$, design box $\mathcal{X}$,
         budget $B$, mini-batch $K$, finite-difference step $\epsilon$, warm-start
         $(N_{\mathrm{ws}}, n_{\mathrm{mc}}^{\text{ws}})$, Adam parameters
         $(\eta, \beta_1, \beta_2)$
\State $\mat{x}_0 \gets$ random-search warm start: best of $N_{\mathrm{ws}}$ designs,
       each scored with $n_{\mathrm{mc}}^{\text{ws}}$ MC samples
\State $b \gets N_{\mathrm{ws}} n_{\mathrm{mc}}^{\text{ws}}$;\quad $t \gets 0$
\While{$b + 2K \leq B$}
  \State draw $\{\boldsymbol{\xi}^{(i)}\}_{i=1}^{K} \sim \rho$ and
         $\{\mat{v}^{(i)}\}_{i=1}^{K} \sim \mathcal{N}(0, I_n)$
  \State \label{step:zo-grad}
         form the two-point estimate $\hat{\mat{g}}^{\mathrm{MC}}_t$ via~\eqref{eq:opt1}
  \State $\mat{x}_{t+1} \gets \Pi_{\mathcal{X}}\!\big(
         \mathrm{Adam}(\mat{x}_t, \hat{\mat{g}}^{\mathrm{MC}}_t;\, \eta, \beta_1,
         \beta_2)\big)$
  \State $b \gets b + 2K$;\quad $t \gets t + 1$
\EndWhile
\State \textbf{return} the final iterate $\mat{x}_t$
\end{algorithmic}
\end{algorithm}

\section{Theoretical Analysis}
\label{sec:theory}

The estimator in~\eqref{eq:opt1} turns a small number of paired
black-box evaluations into a stochastic descent direction. Its
usefulness depends on three questions: which gradient it recovers on
average, how much variation remains in that estimate, and how these
errors translate into simulation cost. This section addresses these
questions in turn.

Because the random perturbations smooth the objective locally, the
natural reference for the zeroth-order estimate is the Gaussian-smoothed
objective
\begin{equation*}
F_\epsilon(\mat{x})
=
\mathbb{E}_{\mat{v}\sim\mathcal{N}(0,I_n)}
\big[F(\mat{x}+\epsilon\mat{v})\big].
\end{equation*}
The analysis uses the following standard regularity conditions; proofs
are deferred to Appendix~\ref{app:proofs}.

\begin{assumption}\label{asm:smooth}
For almost every $\boldsymbol{\xi}$, the per-sample loss
$\ell(\cdot,\boldsymbol{\xi})$ has an $L$-Lipschitz gradient and an
$L_2$-Lipschitz Hessian on $\mathbb{R}^n$, and its gradient is
square-integrable, with per-sample gradient variance
\begin{equation*}
\sigma_g^2(\mat{x})
:=
\mathbb{E}_{\boldsymbol{\xi}}
\big\|
\nabla_{\mat{x}}\ell(\mat{x},\boldsymbol{\xi})
-
\nabla F(\mat{x})
\big\|^2
<\infty.
\end{equation*}
\end{assumption}

The softplus in~\eqref{eq:smoothed-spec} is $C^\infty$, so the required
smoothness is inherited from the simulated circuit metrics. These are
standard assumptions in zeroth-order analysis
\cite{nesterov2017random,ghadimi2013stochastic,berahas2022theoretical}.
The domain is taken as $\mathbb{R}^n$ because the Gaussian perturbations
used in $F_\epsilon$ are unbounded.

\subsection{Estimator Guarantees and Sample Complexity}
\label{sec:theory:results}

\begin{proposition}[Unbiasedness and smoothing bias]\label{prop:bias}
Under Assumption~\ref{asm:smooth}, at any $\mat{x}\in\mathcal{X}$,
\begin{equation}
\mathbb{E}\big[\hat{\mat{g}}^{\mathrm{MC}}\,\big|\,\mat{x}\big]
  = \nabla F_\epsilon(\mat{x}),
\qquad
\big\|\nabla F_\epsilon(\mat{x})-\nabla F(\mat{x})\big\|
  \le \tfrac{1}{2}L_2\,n\,\epsilon^2 .
\label{eq:prop-bias}
\end{equation}
\end{proposition}

The estimator therefore recovers the gradient of the smoothed objective
exactly in expectation, with an $O(\epsilon^2)$ discrepancy from the
original objective. Its usefulness also depends on how much a finite
Monte Carlo estimate fluctuates around this mean.

\begin{proposition}[Process-dimension-free variance]\label{prop:var}
Under Assumption~\ref{asm:smooth} with per-pair common random
numbers,
\begin{equation}
\begin{aligned}
\mathbb{E}\big\|\hat{\mat{g}}^{\mathrm{MC}}-\nabla F_\epsilon(\mat{x})\big\|^2
\le \frac{1}{K}\Big[&2(n{+}4)\|\nabla F(\mat{x})\|^2\\
  &+ (n{+}4)\,\sigma_g^2(\mat{x})\\
  &+ \tfrac{1}{2}L^2(n{+}6)^4\epsilon^2\Big].
\end{aligned}
\label{eq:prop-var}
\end{equation}
\end{proposition}

The variance decreases as $1/K$ and introduces no explicit factor in
the process dimension $d_\xi$; process variation enters through
$\sigma_g^2(\mat{x})$. Sharing the process realization within each pair
also prevents process noise from being amplified as $\epsilon$ becomes
small. Together, Propositions~\ref{prop:bias} and~\ref{prop:var}
characterize the error in the descent direction used by ZO-MC-SGD.
Propagating these errors through the projected update gives the
simulation budget required to approach stationarity.

\begin{theorem}[Sample complexity]\label{thm:rate}
Let Assumption~\ref{asm:smooth} hold with $\|\nabla F(\mat{x})\|\le G$ on
$\mathcal{X}$, and run ZO-MC-SGD~\eqref{eq:descent} for $T$ steps
from $\mat{x}_0$ with constant step $\eta\le 1/(2L)$. Then, for the projected
gradient mapping
$\mathcal{G}_\eta(\mat{x}) = \tfrac{1}{\eta}\big(\mat{x} - \Pi_{\mathcal{X}}(\mat{x} - \eta\nabla F(\mat{x}))\big)$,
whose norm measures stationarity on the design box $\mathcal{X}$,
\begin{equation}
\begin{aligned}
\frac{1}{T}\sum_{t=0}^{T-1}\mathbb{E}\big\|\mathcal{G}_\eta(\mat{x}_t)\big\|^2
={}& O\!\left(\frac{F(\mat{x}_0)-F^\star}{\eta T}
  + \frac{L\eta}{K}\,V_n\right.\\
  &\left.{}+ L_2^2 n^2\epsilon^4\right),
\end{aligned}
\label{eq:thm-rate}
\end{equation}
where
\begin{equation*}
\begin{aligned}
V_n
&:= 2(n{+}4)G^2 + (n{+}4)\bar{\sigma}_g^2
   + \tfrac{1}{2}L^2(n{+}6)^4\epsilon^2,\\
\bar{\sigma}_g^2
&:= \textstyle\sup_{\mat{x}\in\mathcal{X}}\sigma_g^2(\mat{x}),
\qquad
F^\star := \textstyle\inf_{\mat{x}\in\mathcal{X}} F(\mat{x}).
\end{aligned}
\end{equation*}
Choosing
$\eta = \Theta(\sqrt{K/(nT)})$ and $\epsilon$ small enough that
$L_2^2 n^2\epsilon^4\le\nu/3$ and the $\epsilon^2$ term of $V_n$ is
at most $L^2 n$, ZO-MC-SGD reaches
$\frac{1}{T}\sum_t\mathbb{E}\|\mathcal{G}_\eta(\mat{x}_t)\|^2\le\nu$ within
\begin{equation}
B \;=\; O\!\big(n/\nu^2\big) \quad\text{SPICE evaluations,}
\label{eq:thm-budget}
\end{equation}
independent of the mini-batch size $K$ and of the process dimension
$d_\xi$.
\end{theorem}

Thus, reaching $\nu$-stationarity requires
$O(n/\nu^2)$ simulator evaluations, with no explicit dependence on
$d_\xi$. This guarantee concerns the surrogate objective $F$; its link
to yield is established by the rank-alignment procedure in
Section~\ref{sec:method:alpha}.

\subsection{Numerical Validation of the Estimator}
\label{sec:exp:synthetic}
\label{sec:benchmarks:synthetic}

\begin{figure}[t]
\centering
\includegraphics[width=0.99\columnwidth]{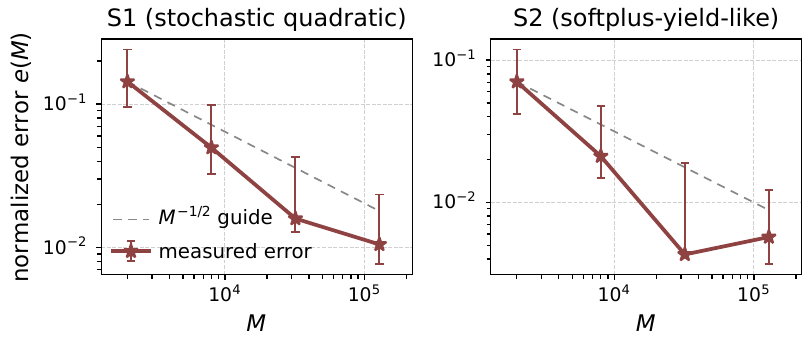}
\caption{Gradient-estimator validation on the synthetic problems.
Normalized error of the mean ZO-MC-SGD estimate versus the number of
independent estimates $M$, with 95\% bootstrap intervals and an
$M^{-1/2}$ reference.}
\Description{Two log-log plots show normalized gradient-estimation
error versus the number of independent estimates for a stochastic
quadratic problem and a softplus yield-like problem. The measured error
generally decreases at the inverse-square-root sampling rate.}
\label{fig:synthetic}
\end{figure}

The preceding analysis describes the estimator in expectation. Before
turning to circuit benchmarks, we compare its implementation with
reference gradients on two synthetic problems where those references
can be computed accurately. S1 provides an exact gradient, while S2
uses the same softplus-based construction as the circuit surrogate and
also admits an analytic yield expression.

\textbf{S1 (stochastic quadratic).}
Consider the per-sample loss
\begin{equation}
\ell(\mat{x}, \boldsymbol{\xi})
=
\tfrac{1}{2}
\|\mat{x}-\boldsymbol{\mu}-\mat{A}\boldsymbol{\xi}\|_2^2,
\qquad
\boldsymbol{\xi}\sim\mathcal{N}(0,I_{d_\xi}),
\label{eq:s1-loss}
\end{equation}
where $\boldsymbol{\mu}\in\mathbb{R}^n$ and
$\mat{A}\in\mathbb{R}^{n\times d_\xi}$. Its expected objective and
gradient are available in closed form,
\begin{equation}
\mathbb{E}_{\boldsymbol{\xi}}[\ell](\mat{x})
=
\tfrac{1}{2}\|\mat{x}-\boldsymbol{\mu}\|_2^2
+
\tfrac{1}{2}\mathrm{tr}(\mat{A}^{\top}\mat{A}),
\qquad
\nabla\mathbb{E}_{\boldsymbol{\xi}}[\ell](\mat{x})
=
\mat{x}-\boldsymbol{\mu}.
\label{eq:s1-grad}
\end{equation}

\textbf{S2 (softplus-yield-like).}
To test the estimator on a non-polynomial surrogate closer to the
circuit setting, S2 uses two affine specifications,
\begin{equation}
g_k(\mat{x},\boldsymbol{\xi})
=
\mat{a}_k^\top\mat{x}
+
\mat{b}_k^\top\boldsymbol{\xi}
+
c_k,
\qquad k=1,2,
\label{eq:s2-specs}
\end{equation}
with yield
\begin{equation}
Y(\mat{x})
=
\mathbb{P}_{\boldsymbol{\xi}}
\!\left(
g_1(\mat{x},\boldsymbol{\xi})\geq0,\;
g_2(\mat{x},\boldsymbol{\xi})\geq0
\right),
\label{eq:s2-yield}
\end{equation}
which has a closed-form expression in terms of the bivariate Gaussian
cumulative distribution function. Applying the
loss construction of Section~\ref{sec:method:loss} produces a
non-polynomial objective. Its reference gradient is computed with
$2\times10^6$ Monte Carlo samples, with a self-consistency error below
$0.2\%$. The analytic yield in~\eqref{eq:s2-yield} is also used for the
rank-alignment test of Section~\ref{sec:method:alpha}.

We evaluate both problems at $n=6$ and $d_\xi=10$ using the deployed
settings $K=4$ and $\epsilon=5\times10^{-3}$. Fig.~\ref{fig:synthetic}
plots the normalized error of the averaged estimate,
\[
e(M)
=
\frac{
\left\|
M^{-1}\sum_m\hat{\mat{g}}_m
-
\nabla\mathbb{E}_{\boldsymbol{\xi}}[\ell]
\right\|
}{
\left\|
\nabla\mathbb{E}_{\boldsymbol{\xi}}[\ell]
\right\|
},
\]
against the number of independent gradient estimates $M$.

On both problems, the error decreases at the $M^{-1/2}$ sampling rate
shown in Fig.~\ref{fig:synthetic}; on S2 it reaches $0.57\%$ at
$M=1.28\times10^5$. Varying $\epsilon$ over a fourfold range produces
no visible bias floor, consistent with the $O(\epsilon^2)$ smoothing
error predicted by Proposition~\ref{prop:bias}.

\section{Experimental Setup}
\label{sec:benchmarks}

We evaluate ZO-MC-SGD on five SPICE-simulated analog circuits with
design dimensions from 6 to 30 and process dimensions from 10 to 42.
The comparison includes five black-box and learning-based baselines
under the same SPICE optimization budgets. This section describes the circuit
benchmarks, the objective used by each method, and the common
evaluation protocol; implementation details are collected in
Appendix~\ref{app:config}.

\subsection{Circuit Benchmarks}
\label{sec:benchmarks:selection}
\label{sec:benchmarks:circuits}

\begin{paperwidefigure}[!t]
\centering
\subfloat[1-stage csamp]{%
  \includegraphics[width=0.36\linewidth]{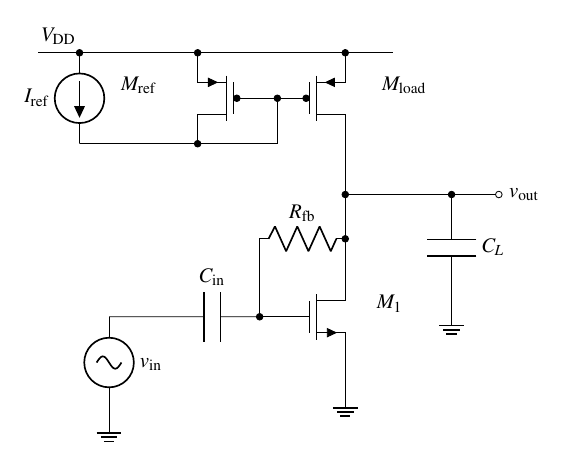}}\hfill
\subfloat[2-stage csmiller]{%
  \includegraphics[width=0.54\linewidth]{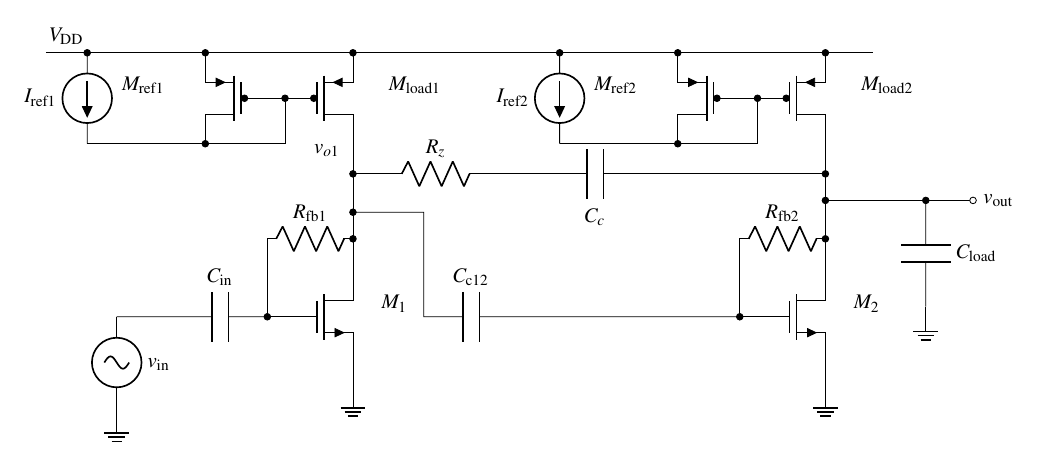}}
\caption{Base circuit of each family: (a)~a common-source stage and
(b)~a two-stage Miller amplifier. Larger benchmarks add stages;
counts are listed in Table~\ref{tab:circuits}.}
\Description{Two circuit schematics. The left schematic is a one-stage
common-source amplifier with an input coupling capacitor, bias resistor,
current-mirror load, and load capacitor. The right schematic is a
two-stage Miller-compensated amplifier with two biased common-source
stages, a compensation resistor and capacitor, and a load capacitor.}
\label{fig:circuit_roster}
\end{paperwidefigure}

Each benchmark is screened before optimization to ensure that it
represents a meaningful yield-design problem. The nominal reference
design must meet all specifications, every specification must be
sensitive to process variation, and its yield must lie between $0.30$
and $0.85$ so that there is substantial room for optimization. The
calibrated loss of Section~\ref{sec:method:loss} must also satisfy the
rank-alignment criterion $\rho_s \leq -0.7$. Each rank-alignment test
uses $N_{\mathrm{cal}}=81$ perturbed calibration designs.

The five benchmarks comprise two topology families: common-source (CS)
cascades with one, three, and five stages, and Miller-compensated
amplifiers with two and three stages.
Adding stages increases both the design dimension $n$ and the process
dimension $d_\xi$, providing a controlled progression in problem size
within each family. Table~\ref{tab:circuits} summarizes the five
benchmarks, and Fig.~\ref{fig:circuit_roster} shows the base topology
of each family.

All benchmarks are simulated in ngspice using LEVEL-1
(Shichman--Hodges) device models~\cite{shichman1968modeling}, with
independent zero-mean Gaussian device mismatch following Pelgrom
scaling~\cite{pelgrom1989}. Under each process realization, a sample is
counted as passing only when gain, unity-gain bandwidth, and static
power simultaneously satisfy the circuit-specific limits in
Table~\ref{tab:specs}. Device parameters and per-benchmark variation
scales are given in Appendix~\ref{app:config}.

The nominal reference design is used to screen and calibrate each
benchmark. Optimization starts from a separate design shared by all
methods; its yield $y_{\mathrm{init}}$ is reported in
Table~\ref{tab:circuits}.

\begin{table}[t]
\centering
\caption{Per-circuit yield specifications ($V_{DD}=1$\,V).}
\label{tab:specs}
\footnotesize
\setlength{\tabcolsep}{6pt}
\begin{tabular}{l c c c}
\toprule
Circuit & Gain (dB) & UGBW (MHz) & Power (mW) \\
\midrule
1-stage csamp     & $\geq 20$ & $\geq 1.0$  & $\leq 0.5$ \\
3-stage csamp     & $\geq 60$ & $\geq 50$   & $\leq 0.5$ \\
5-stage csamp     & $\geq 95$ & $\geq 30$   & $\leq 0.7$ \\
2-stage csmiller  & $\geq 50$ & $\geq 0.2$  & $\leq 0.5$ \\
3-stage csmiller  & $\geq 60$ & $\geq 0.05$ & $\leq 0.7$ \\
\bottomrule
\end{tabular}
\end{table}

\begin{table}[t]
\centering
\caption{Circuit benchmarks. csamp denotes a common-source cascade and
csmiller a Miller-compensated cascade.}
\label{tab:circuits}
\footnotesize
\setlength{\tabcolsep}{4pt}
\begin{tabular}{l c c c c c}
\toprule
Circuit & $n$ & $d_\xi$ & $y_{\mathrm{init}}$
        & $\sigma$-scale & $\rho_s$ \\
\midrule
1-stage csamp     &  6 & 10 & 0.24 & 3.0 & $-0.977$ \\
3-stage csamp     & 18 & 26 & 0.20 & 1.5 & $-0.873$ \\
5-stage csamp     & 30 & 42 & 0.34 & 0.1 & $-0.915$ \\
2-stage csmiller  & 14 & 26 & 0.64 & 0.1 & $-0.876$ \\
3-stage csmiller  & 22 & 38 & 0.66 & 0.1 & $-0.953$ \\
\bottomrule
\end{tabular}
\end{table}

\subsection{Baselines and Objectives}
\label{sec:benchmarks:baselines}

We compare ZO-MC-SGD with five baselines: Bayesian optimization
(BO)~\cite{jones1998efficient,snoek2012practical,frazier2018tutorial},
the covariance matrix adaptation evolution strategy
(CMA-ES)~\cite{hansen2001completely,hansen2016cma}, particle swarm
optimization (PSO)~\cite{kennedy1995particle}, trust-region Bayesian
optimization (TuRBO)~\cite{eriksson2019turbo}, and the variation-aware
reinforcement-learning method
RobustAnalog~\cite{shi2022robustanalog}.

Each method is run on the objective suited to its optimization
mechanism. ZO-MC-SGD uses the smooth loss of
Section~\ref{sec:method:loss}, since its local finite-difference update
requires a continuous optimization signal. BO, CMA-ES, PSO, and TuRBO
can instead optimize the direct Monte Carlo yield estimate
$\hat{Y}(\mat{x})$, while RobustAnalog uses its specification-margin
reward. Section~\ref{sec:exp:objalign} shows that direct yield gives
the four black-box baselines better final designs than the smooth loss.

Because these objectives require different numbers of SPICE simulations
per algorithmic query or update, all methods are compared under the
same per-run SPICE optimization budget rather than the same number of
optimizer iterations. Table~\ref{tab:perquery} summarizes these costs.

\begin{table}[t]
\centering
\caption{SPICE cost of each algorithmic query or update.}
\label{tab:perquery}
\footnotesize
\ifdefined\ARXIVVERSION
  \setlength{\tabcolsep}{3pt}
\else
  \setlength{\tabcolsep}{6pt}
\fi
\renewcommand{\arraystretch}{1.15}
\begin{tabular}{l l l}
\toprule
\rowcolor{tabBandYield}
\textbf{Method} & \textbf{SPICE cost} & \textbf{Estimated quantity} \\
\midrule
ZO-MC-SGD (proposed)
  & $8$/step
  & ZO gradient estimate $\hat{\mat{g}}$ \\
\rowcolor{tabRowAlt}
BO, CMA-ES, PSO, TuRBO
  & $10$/query
  & yield $\hat{Y}(\mat{x})$ \\
RobustAnalog
  & $20$/step
  & specification-margin reward \\
\bottomrule
\end{tabular}
\end{table}

\subsection{Evaluation Protocol}
\label{sec:benchmarks:protocol}

Each method is run on every circuit at SPICE budgets
$B\in\{25,50,100,200,400,800,1600,3200\}$ and repeated with five
random seeds. The yield of every returned design is re-estimated on the
same held-out set of
$n_{\mathrm{mc}}^{\mathrm{eval}}=80$ process samples, disjoint from
those used during optimization. We report the mean and observed range over the five runs.

Our primary sample-efficiency metric is the minimum SPICE budget at
which the mean yield first reaches $Y^\star=0.95$, a high-yield target
used in design-for-yield studies
\cite{wang2022freezethaw,guo2023bnn}. Budgets up to $B=3200$ are
included to track methods that require substantially more simulation
to reach the same target.

The budget $B$ counts SPICE evaluations within each optimization run,
including the random-search warm start of ZO-MC-SGD; the one-time
rank-alignment calibration is treated separately. As shown in the
ablation of Section~\ref{sec:exp:objalign}, using a single shared
sharpness value leaves the target-reaching budgets unchanged across all
five circuits. The remaining optimization settings are shared across
benchmarks and are listed in Appendix~\ref{app:config}.

\ifdefined\ARXIVVERSION
\FloatBarrier
\fi

\section{Results}
\label{sec:experiments}

We next examine whether the stochastic descent directions developed
above translate into lower simulation cost for circuit-level yield
optimization.

\subsection{Sample Efficiency}
\label{sec:exp:performance}

Fig.~\ref{fig:conv} shows how the mean yield changes with SPICE budget, while Table~\ref{tab:minbudget} reports the minimum budget at which each
method reaches the target mean yield $Y^\star=0.95$. Under the common
evaluation protocol, ZO-MC-SGD reaches this target within 50--200
simulations on all five benchmarks. It ties the best baseline on
1-stage csamp and 2-stage csmiller; on the other three circuits, the
best baseline requires four, eight, and four times as many simulations,
respectively. Even at $B=3200$, CMA-ES does not reach the target on
3-stage csamp, while RobustAnalog does not reach it on three of the five
circuits.

\begin{table}[t]
\centering
\caption{Minimum SPICE budget at which the five-seed mean yield reaches
0.95 under the common evaluation protocol. $>3200$ indicates that the
target was not reached; RA = RobustAnalog.}
\label{tab:minbudget}
\ifdefined\ARXIVVERSION
  \scriptsize
  \setlength{\tabcolsep}{1.5pt}
\else
  \footnotesize
  \setlength{\tabcolsep}{6pt}
\fi
\renewcommand{\arraystretch}{1.25}
\begin{tabular}{l c c c c c c}
\toprule
\rowcolor{tabBandYield}
\textbf{Circuit ($d_\xi$)} & \textbf{ZO-MC-SGD} & \textbf{BO} &
\textbf{CMA-ES} & \textbf{PSO} & \textbf{TuRBO} & \textbf{RA} \\
\midrule
1-stage csamp (10) & \textbf{200} & \textbf{200} & 800 & 400 & 400 & 1600 \\
\rowcolor{tabRowAlt}
3-stage csamp (26) & \textbf{200} & 1600 & $>3200$ & 800 & 800 & $>3200$ \\
5-stage csamp (42) & \textbf{100} & 3200 & 800 & 1600 & 800 & $>3200$ \\
\rowcolor{tabRowAlt}
2-stage csmiller (26) & \textbf{100} & 400 & \textbf{100} & 400 & 400 & $>3200$ \\
3-stage csmiller (38) & \textbf{50} & 200 & 400 & 400 & 800 & 1600 \\
\bottomrule
\end{tabular}
\end{table}

\begin{paperwidefigure}[!t]
\centering
\ifdefined\ARXIVVERSION
\subfloat[CS cascade family\label{fig:conv_csamp}]{%
  \includegraphics[width=0.96\textwidth]{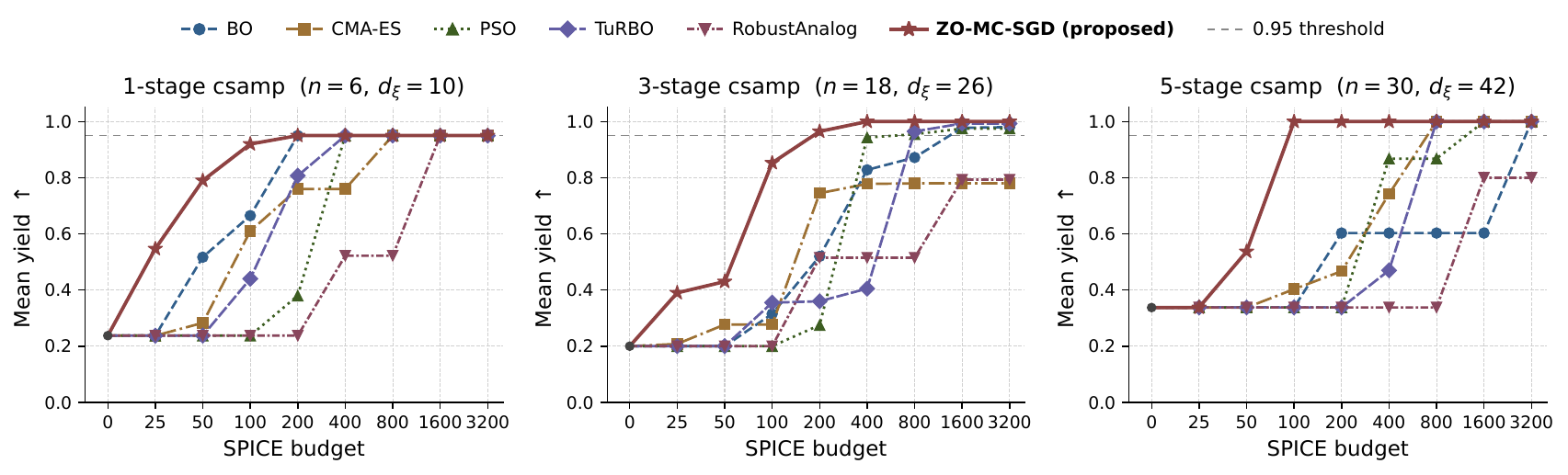}}\\[6pt]
\subfloat[Miller-compensated family\label{fig:conv_miller}]{%
  \includegraphics[width=0.72\textwidth]{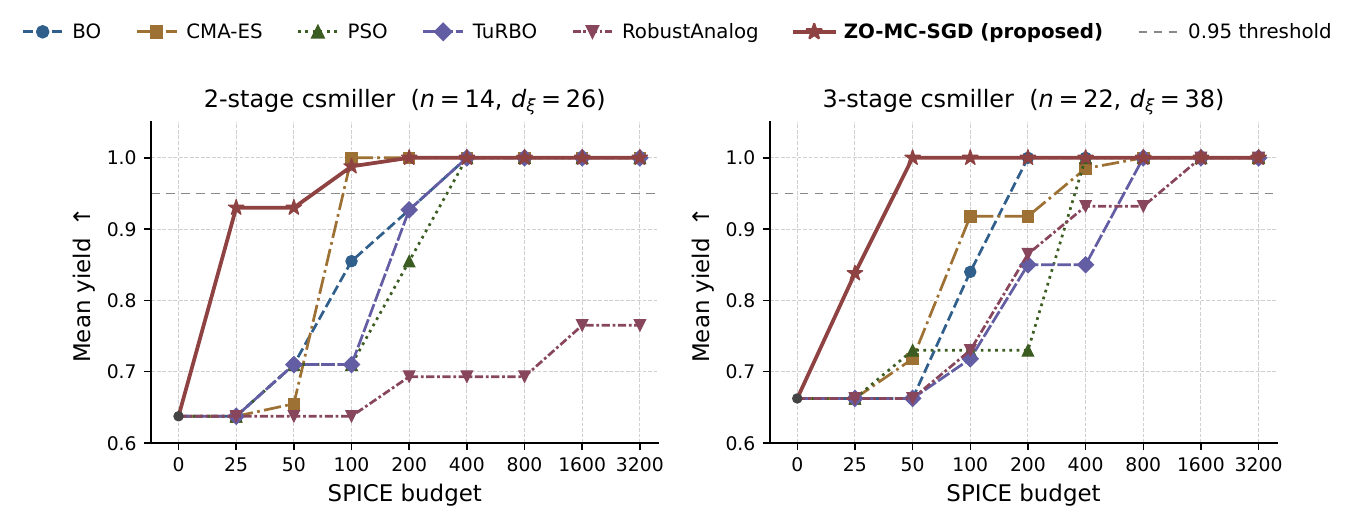}}
\else
\begin{subfigure}{\linewidth}
  \centering
  \includegraphics[width=\linewidth]{figures/fig_yield_vs_budget_cs_amp.pdf}
  \caption{CS cascade family}
  \label{fig:conv_csamp}
\end{subfigure}\\[6pt]
\begin{subfigure}{\linewidth}
  \centering
  \includegraphics[width=0.755\linewidth]{figures/fig_yield_vs_budget_cs_se_miller.pdf}
  \caption{Miller-compensated family}
  \label{fig:conv_miller}
\end{subfigure}
\fi
\caption{Mean yield versus SPICE budget for (a)~the CS cascade family
and (b)~the Miller-compensated family. The dashed line marks the 0.95
yield target.}
\Description{Five line plots compare mean yield as the SPICE budget
increases for one-, three-, and five-stage common-source cascades and
two- and three-stage Miller-compensated amplifiers. ZO-MC-SGD reaches
the 0.95 threshold within 50 to 200 simulations on every benchmark,
with the clearest separation from the baselines on the larger circuits.}
\label{fig:conv}
\end{paperwidefigure}

The separation is clearest on the larger circuits, especially the
3- and 5-stage CS cascades. ZO-MC-SGD reaches the target with 200 and
100 simulations, respectively, while the strongest baselines require
800 simulations on both circuits; the remaining methods require
1600--3200 simulations or do not reach the target within the tested
range.

This pattern reflects how the methods use the early simulation budget.
The competing methods must first build a model, evaluate a population,
or train a policy before obtaining useful design updates. After its
short warm start, ZO-MC-SGD instead turns each $2K=8$-simulation batch
directly into a descent direction. The asymmetry is in what a simulation
buys: the four direct-yield black-box baselines spend ten simulations to
estimate one scalar yield at a candidate design, whereas ZO-MC-SGD
spends eight to estimate a direction in design space. The resulting
budget gap widens with stage count in both circuit families.

The target crossings in Table~\ref{tab:minbudget} are based on the
common 80-sample evaluation protocol. To estimate the absolute yield
more precisely, we re-evaluate the ZO-MC-SGD designs returned at these
budgets using 1000 fresh process samples
(Table~\ref{tab:rescore}). The five-seed mean remains above 0.95 on
four circuits, at 0.976, 1.000, 0.976, and 1.000. On 1-stage csamp, the
high-precision estimate is 0.934, with a 95\% interval of
$[0.917,0.949]$. No method exceeds 0.95 under the common protocol on
this benchmark at any tested budget, suggesting an empirical yield
ceiling within the tested design space.

\begin{table}[t]
\centering
\caption{High-precision yield re-estimates of the ZO-MC-SGD designs
returned at the target budgets in Table~\ref{tab:minbudget}. Each
design is re-estimated with $N=1000$ fresh process samples; entries
report the five-seed mean and observed range.}
\label{tab:rescore}
\footnotesize
\setlength{\tabcolsep}{6pt}
\renewcommand{\arraystretch}{1.15}
\begin{tabular}{l c c}
\toprule
\rowcolor{tabBandYield}
\textbf{Circuit ($B$)} & \textbf{Mean yield} & \textbf{Per-seed range} \\
\midrule
1-stage csamp (200) & $0.934$ & $[0.934, 0.934]$ \\
\rowcolor{tabRowAlt}
3-stage csamp (200) & $0.976$ & $[0.973, 0.985]$ \\
5-stage csamp (100) & $1.000$ & $[1.000, 1.000]$ \\
\rowcolor{tabRowAlt}
2-stage csmiller (100) & $0.976$ & $[0.878, 1.000]$ \\
3-stage csmiller (50) & $1.000$ & $[1.000, 1.000]$ \\
\bottomrule
\end{tabular}
\end{table}

\subsection{Ablations and Robustness}
\label{sec:exp:objalign}

To separate the effect of objective choice from that of the optimizer,
we rerun BO, CMA-ES, PSO, and TuRBO on the softplus surrogate with the
same optimizer settings, budgets, seeds, and evaluation protocol.
Table~\ref{tab:objablation} reports the resulting mean-yield difference,
defined as direct Monte Carlo yield minus the softplus surrogate. Direct
yield performs better on every circuit for all four methods; averaged
over the five circuits, the gain ranges from $0.24$ to $0.41$. The main
comparison therefore evaluates these baselines on direct yield, while
the softplus surrogate serves a different role in ZO-MC-SGD by providing
an informative local finite-difference signal. RobustAnalog is not
included because it optimizes its own specification-margin reward.

\begin{table}[t]
\centering
\caption{Baseline objective alignment. Mean-yield difference between
direct Monte Carlo yield and the softplus surrogate; positive values
favor direct yield.}
\label{tab:objablation}
\footnotesize
\setlength{\tabcolsep}{6pt}
\renewcommand{\arraystretch}{1.15}
\begin{tabular}{l c c c c}
\toprule
\rowcolor{tabBandYield}
\textbf{Circuit ($d_\xi$)} & \textbf{BO} & \textbf{CMA-ES} &
\textbf{PSO} & \textbf{TuRBO} \\
\midrule
1-stage csamp (10)    & $+0.25$ & $+0.13$ & $+0.16$ & $+0.34$ \\
\rowcolor{tabRowAlt}
3-stage csamp (26)    & $+0.29$ & $+0.14$ & $+0.32$ & $+0.19$ \\
5-stage csamp (42)    & $+0.39$ & $+0.27$ & $+0.44$ & $+0.36$ \\
\rowcolor{tabRowAlt}
2-stage csmiller (26) & $+0.35$ & $+0.40$ & $+0.37$ & $+0.55$ \\
3-stage csmiller (38) & $+0.58$ & $+0.27$ & $+0.56$ & $+0.62$ \\
\midrule
All circuits          & $+0.37$ & $+0.24$ & $+0.37$ & $+0.41$ \\
\bottomrule
\end{tabular}
\end{table}

\phantomsection\label{sec:exp:alpha}
We next examine sensitivity to the softplus sharpness. Five settings
per circuit span a $50\times$ range, and all 25 circuit--setting
combinations satisfy the rank-alignment criterion
$\rho_s\leq-0.7$. Replacing the circuit-specific settings with a single
shared default leaves the target-reaching budget unchanged on every
circuit, while the average yield difference between the two
configurations is below $0.001$. Circuit-specific calibration therefore
provides no measurable sample-efficiency advantage on these benchmarks.

Finally, we test whether the sample-efficiency advantage persists under
correlated process variation. On the two circuits with $d_\xi=26$, we
replace the independent process model with
$\boldsymbol{\xi}\sim\mathcal{N}(\mathbf{0},DCD)$, where $D$ contains
the calibrated marginal standard deviations and $C$ assigns a
correlation of $0.5$ to parameters sharing oxide, doping, or
lithographic effects and to corresponding device parameters within each
stage. This change shifts the initial yield from $0.219$ to $0.172$ on
3-stage csamp and from $0.641$ to $0.750$ on 2-stage csmiller.
ZO-MC-SGD still reaches $Y^\star$ with 200 simulations on 3-stage
csamp, compared with 400 for CMA-ES and more than 400 for BO; on
2-stage csmiller, the corresponding budgets are 25, 100, and 200.
Thus, the estimator does not require independent process components,
only independent samples from their joint distribution.

\ifdefined\ARXIVVERSION
\FloatBarrier
\fi

\section{Discussion}
\label{sec:discussion}

The results place ZO-MC-SGD in a clear operating regime:
yield-optimization problems in which SPICE simulations are expensive
and the available budget is small relative to the design dimension. In
this regime, a small batch of simulations can be turned directly into a
local design update, without first fitting a global model, evaluating a
population, or training a policy. The growing advantage on the larger
benchmarks is consistent with this distinction becoming more important
as the design dimension increases.

The main modeling requirement is access to a process distribution
$\rho(\boldsymbol{\xi})$ from which samples can be drawn. If this
distribution is itself uncertain or changes across manufacturing
conditions, the problem becomes one of optimization under
distributional uncertainty rather than yield optimization under a
fixed process model. Distributionally robust formulations provide a
natural extension to that setting~\cite{pan2023distributionally}.

The theoretical bounds contain no explicit dependence on the process
dimension, but the experiments cover only up to $d_\xi=42$. Whether
the same behavior persists in production-scale circuits with hundreds
of mismatch variables remains to be established. Extending the method
to this regime may require variance reduction or structure-exploiting
zeroth-order estimators.

\section{Conclusion}
\label{sec:conclusion}

We introduced ZO-MC-SGD, which turns specification margins from
black-box SPICE simulations into stochastic zeroth-order descent
directions for yield optimization. Across five analog circuit
benchmarks, it reaches a mean yield of $0.95$ on four circuits within
50--200 simulations and the empirical yield ceiling on the fifth,
reducing the required simulation budget by up to a factor of eight
relative to the best baseline. These results show the value of
stochastic zeroth-order updates when simulation budgets are tight and
the process distribution is available for sampling.


\ifdefined\ARXIVVERSION
\appendices
\else
\appendix
\fi

\section{Proofs of Section~\protect\ref{sec:theory}}
\label{app:proofs}

We use the Gaussian-smoothing identity of Nesterov and
Spokoiny~\cite{nesterov2017random}: for differentiable $h$ with
$h_\epsilon(\mat{x})=\mathbb{E}_{v\sim\mathcal{N}(0,I_n)}[h(\mat{x}+\epsilon \mat{v})]$,
\begin{equation}
\begin{aligned}
\nabla h_\epsilon(\mat{x})
  &= \tfrac{1}{\epsilon}\,\mathbb{E}_v\!\big[
     h(\mat{x}+\epsilon \mat{v})\,\mat{v}\big]\\
  &= \tfrac{1}{2\epsilon}\,\mathbb{E}_v\!\big[
     (h(\mat{x}+\epsilon \mat{v})
     -h(\mat{x}-\epsilon \mat{v}))\,\mat{v}\big],
\end{aligned}
\label{eq:gs-identity}
\end{equation}
the second equality following from the symmetry $\mat{v}\mapsto-\mat{v}$ of the
Gaussian.

\subsection*{Proof of Proposition~\ref{prop:bias}}
\emph{Unbiasedness.} Condition on $\mat{x}$ and fix one pair $(\boldsymbol{\xi},\mat{v})$.
With $\boldsymbol{\xi}$ held fixed, applying~\eqref{eq:gs-identity} to
$h=\ell(\cdot,\boldsymbol{\xi})$ gives
\begin{equation}
\mathbb{E}_v\!\Big[\tfrac{1}{2\epsilon}
  \big(\ell(\mat{x}{+}\epsilon \mat{v},\boldsymbol{\xi})-\ell(\mat{x}{-}\epsilon \mat{v},\boldsymbol{\xi})\big)\mat{v}\Big]
  = \nabla_x\ell_\epsilon(\mat{x},\boldsymbol{\xi}).
\end{equation}
Taking the expectation over $\boldsymbol{\xi}\sim\rho$ and exchanging it with the
gradient (justified by Assumption~\ref{asm:smooth} and dominated
convergence),
\begin{equation}
\mathbb{E}_\xi\big[\nabla_x\ell_\epsilon(\mat{x},\boldsymbol{\xi})\big]
  = \nabla F_\epsilon(\mat{x}).
\end{equation}
The $K$ summands of $\hat{\mat{g}}^{\mathrm{MC}}$ are i.i.d.\ with this mean,
so $\mathbb{E}[\hat{\mat{g}}^{\mathrm{MC}}\,|\,\mat{x}]=\nabla F_\epsilon(\mat{x})$.

\emph{Smoothing bias.} Since $F=\mathbb{E}_\xi[\ell]$ inherits the
$L_2$-Lipschitz Hessian, the gradient Taylor remainder obeys
\begin{equation}
\big\|\nabla F(\mat{x}{+}\epsilon \mat{v})-\nabla F(\mat{x})-\epsilon\nabla^2F(\mat{x})\mat{v}\big\|
  \le \tfrac{L_2}{2}\epsilon^2\|\mat{v}\|^2 .
\end{equation}
Using $\nabla F_\epsilon(\mat{x})=\mathbb{E}_v[\nabla F(\mat{x}+\epsilon \mat{v})]$ and
$\mathbb{E}_v[\mat{v}]=0$,
\begin{equation}
\big\|\nabla F_\epsilon(\mat{x})-\nabla F(\mat{x})\big\|
  \le \tfrac{L_2}{2}\epsilon^2\,\mathbb{E}_v\|\mat{v}\|^2
  = \tfrac{L_2}{2}\,n\,\epsilon^2 ,
\end{equation}
as $\mathbb{E}_v\|\mat{v}\|^2=n$. \hfill\rule{1.1ex}{1.1ex}

\subsection*{Proof of Proposition~\ref{prop:var}}
The $K$ summands are i.i.d., so
\begin{equation}
\mathbb{E}\big\|\hat{\mat{g}}^{\mathrm{MC}}-\nabla F_\epsilon\big\|^2
  = \tfrac{1}{K}\big(\mathbb{E}\|\mat{g}^{(1)}\|^2
    -\|\nabla F_\epsilon\|^2\big)
  \le \tfrac{1}{K}\mathbb{E}\|\mat{g}^{(1)}\|^2,
\end{equation}
where $\mat{g}^{(1)}=D(\boldsymbol{\xi},\mat{v})\,\mat{v}$ and
\begin{equation}
D = \tfrac{1}{2\epsilon}\big(\ell(\mat{x}{+}\epsilon \mat{v},\boldsymbol{\xi})
  -\ell(\mat{x}{-}\epsilon \mat{v},\boldsymbol{\xi})\big).
\end{equation}
Split $D=D_F(\mat{v})+\Delta(\boldsymbol{\xi},\mat{v})$ with
\begin{equation}
\begin{aligned}
D_F(\mat{v})&=\tfrac{1}{2\epsilon}\big(F(\mat{x}{+}\epsilon \mat{v})
       -F(\mat{x}{-}\epsilon \mat{v})\big),\\
\Delta&=\tfrac{1}{2\epsilon}\big(\tilde\ell(\mat{x}{+}\epsilon \mat{v},\boldsymbol{\xi})
       -\tilde\ell(\mat{x}{-}\epsilon \mat{v},\boldsymbol{\xi})\big),
\end{aligned}
\end{equation}
where $\tilde\ell=\ell-F$. As $\mathbb{E}_\xi[\Delta]=0$ and $D_F$ is
$\boldsymbol{\xi}$-free, the cross term vanishes and
\begin{equation}
\mathbb{E}\|\mat{g}^{(1)}\|^2
  = \mathbb{E}_v\big[(D_F^2+\mathbb{E}_\xi[\Delta^2])\|\mat{v}\|^2\big].
\end{equation}

\emph{Mean term.} $D_F\,\mat{v}$ is the two-point estimator of the
deterministic $F$; the Nesterov--Spokoiny second-moment
bound~\cite{nesterov2017random} gives
\begin{equation}
\mathbb{E}_v\big[D_F^2\|\mat{v}\|^2\big]
  \le 2(n{+}4)\|\nabla F(\mat{x})\|^2
    +\tfrac{1}{2}L^2(n{+}6)^3\epsilon^2 .
\end{equation}

\emph{Process term.} Writing
\begin{equation}
\Delta = \tfrac{1}{2\epsilon}\int_{-\epsilon}^{\epsilon}
  \nabla\tilde\ell(\mat{x}+s\mat{v},\boldsymbol{\xi})^\top \mat{v}\,ds
\end{equation}
and expanding about $s=0$ gives
$\Delta=\nabla\tilde\ell(\mat{x},\boldsymbol{\xi})^\top \mat{v}+r$, with the remainder bounded
by $|r|\le L\epsilon\|\mat{v}\|^2$ (the gradient of $\tilde\ell$
is $2L$-Lipschitz). Splitting
$\Delta^2\le\big(1{+}\tfrac{2}{n+2}\big)(\nabla\tilde\ell^\top \mat{v})^2
+\big(1{+}\tfrac{n+2}{2}\big)r^2$ by Young's inequality and applying
the Gaussian identities
$\mathbb{E}_v[(\mat{v}^\top C \mat{v})\|\mat{v}\|^2]=(n{+}2)\operatorname{tr}C$
and $\mathbb{E}_v\|\mat{v}\|^6=n(n{+}2)(n{+}4)$, with
$C=\operatorname{Cov}_\xi(\nabla\ell(\mat{x},\boldsymbol{\xi}))$ and
$\operatorname{tr}C=\sigma_g^2(\mat{x})$, yields
\begin{equation}
\mathbb{E}_v\big[\mathbb{E}_\xi[\Delta^2]\|\mat{v}\|^2\big]
  \le (n{+}4)\,\sigma_g^2(\mat{x})
  + \tfrac{1}{2}L^2\,n(n{+}2)(n{+}4)^2\,\epsilon^2 .
\end{equation}
The shared $\boldsymbol{\xi}$ keeps
$\Delta=O(\|\nabla\tilde\ell\|)$ rather than $O(1/\epsilon)$. With
independent draws, the numerator instead contains uncancelled loss
noise of scale $\sigma_\ell=(\operatorname{Var}_\xi\,\ell)^{1/2}$,
producing an $O(\sigma_\ell^2/\epsilon^2)$ contribution to the
variance. Summing the two terms, with
$(n{+}6)^3+n(n{+}2)(n{+}4)^2\le(n{+}6)^4$, and dividing by $K$
gives~\eqref{eq:prop-var}.
\hfill\rule{1.1ex}{1.1ex}

\subsection*{Proof of Theorem~\ref{thm:rate}}
We give the unconstrained case ($\Pi_{\mathcal{X}}=\mathrm{id}$,
$\mathcal{G}_\eta=\nabla F$), for which the constants inside the
$O(\cdot)$ of~\eqref{eq:thm-rate} are $(4,\,2,\,\tfrac{3}{2})$; the
projected case replaces $\nabla F$ by the gradient mapping and
follows the mini-batch proximal argument of Ghadimi, Lan, and
Zhang~\cite{ghadimi2016minibatch}, changing only the absolute
constants. By $L$-smoothness,
\begin{equation}
F(\mat{x}_{t+1})\le F(\mat{x}_t)-\eta\langle\nabla F(\mat{x}_t),\hat{\mat{g}}^{\mathrm{MC}}_t\rangle
  +\tfrac{L\eta^2}{2}\|\hat{\mat{g}}^{\mathrm{MC}}_t\|^2 .
\end{equation}
Condition on $\mat{x}_t$, write $\mat{b}_t=\nabla F_\epsilon(\mat{x}_t)-\nabla F(\mat{x}_t)$
with $\|\mat{b}_t\|\le\tfrac{L_2 n}{2}\epsilon^2=:\bar b$
(Proposition~\ref{prop:bias}), and use
$\mathbb{E}[\hat{\mat{g}}^{\mathrm{MC}}_t]=\nabla F_\epsilon(\mat{x}_t)$ and
$\mathbb{E}\|\hat{\mat{g}}^{\mathrm{MC}}_t\|^2\le\|\nabla F_\epsilon(\mat{x}_t)\|^2+V_n/K$
(Proposition~\ref{prop:var}, with $G$ bounding $\|\nabla F\|$):
\begin{equation}
\begin{aligned}
\mathbb{E}[F(\mat{x}_{t+1})]\le{}& F(\mat{x}_t)
 -\eta\langle\nabla F(\mat{x}_t),\nabla F(\mat{x}_t)+\mat{b}_t\rangle\\
 &+\tfrac{L\eta^2}{2}\big(\|\nabla F(\mat{x}_t)+\mat{b}_t\|^2+\tfrac{V_n}{K}\big).
\end{aligned}
\end{equation}
For $\eta\le 1/(2L)$, so that $L\eta^2\le\eta/2$, the Young inequalities
\begin{equation}
\begin{aligned}
\langle\nabla F,\mat{b}_t\rangle&\ge-\tfrac14\|\nabla F\|^2-\|\mat{b}_t\|^2,\\
\|\nabla F+\mat{b}_t\|^2&\le 2\|\nabla F\|^2+2\bar b^2
\end{aligned}
\end{equation}
leave a $-\tfrac{\eta}{4}\|\nabla F(\mat{x}_t)\|^2$ descent term, the
$\|\nabla F\|^2$ coefficient being
$-\tfrac{3}{4}\eta+L\eta^2\le-\tfrac{\eta}{4}$ and the $\bar b^2$
coefficient $\eta+L\eta^2\le\tfrac{3}{2}\eta$:
\begin{equation}
\tfrac{\eta}{4}\mathbb{E}\|\nabla F(\mat{x}_t)\|^2
  \le \mathbb{E}[F(\mat{x}_t)-F(\mat{x}_{t+1})]
  +\tfrac{L\eta^2}{2}\tfrac{V_n}{K}+\tfrac{3}{2}\eta \bar b^2 .
\end{equation}
Summing over $t=0,\dots,T-1$, telescoping, and dividing by
$\tfrac{\eta T}{4}$,
\begin{equation}
\tfrac{1}{T}\sum_{t}\mathbb{E}\|\nabla F(\mat{x}_t)\|^2
  \le \tfrac{4(F(\mat{x}_0)-F^\star)}{\eta T}
    +\tfrac{2L\eta}{K}V_n+6\bar b^2,
\end{equation}
which is~\eqref{eq:thm-rate} with the constants
$(4,\,2,\,\tfrac{3}{2})$, as $6\bar b^2=\tfrac{3}{2}L_2^2n^2\epsilon^4$.
Under the stated $\epsilon$ conditions,
$V_n\le 2(n{+}4)G^2+(n{+}4)\bar\sigma_g^2+L^2 n=O(n)$ and
$6\bar b^2\le\nu/2$; the choice $\eta=\Theta(\sqrt{K/(nT)})$ then
drives the first two terms below $\nu/2$ once $KT=O(n/\nu^2)$, i.e.\
$B=2KT=O(n/\nu^2)$ simulations, independent of
$K$ and $d_\xi$. \hfill\rule{1.1ex}{1.1ex}

\section{Experiment Configuration}
\label{app:config}

\paragraph{Device models and mismatch.}
The LEVEL-1 (Shichman--Hodges) models are chosen for controlled,
reproducible dimension sweeps. Process variation $\boldsymbol{\xi}$
collects per-device shifts in $V_{\mathrm{th}}$, $\beta$, $L$, and $W$
plus two global shifts, so $d_\xi$ grows with stage count.

\paragraph{Benchmark variation scale.}
The $\sigma$-scale of Table~\ref{tab:circuits} is a multiplicative
factor on the per-device baseline mismatch sigmas. Starting from the
baseline sigmas ($\sigma$-scale $=1$), the yield at
$\mat{x}_{\mathrm{NOMINAL}}$ is evaluated with $64$ Monte Carlo samples
on a fixed $\boldsymbol{\xi}$-set, and the scale is swept over a fixed
grid until $y(\mat{x}_{\mathrm{NOMINAL}}) \in [0.30, 0.85]$. The chosen
scale is then fixed for calibration, optimization, and evaluation.

\paragraph{Process sensitivity.}
At the nominal design, each specification responds to at least one
process axis under $\pm\sigma$ perturbations. During optimization, the
full process vector is sampled.

\paragraph{Loss parameters.}
All five benchmarks use
$(w_{\mathrm{gain}},w_{\mathrm{ugbw}},w_{\mathrm{power}})=(1,2,1)$
and $\gamma=0.1$. For the circuit-specific calibration used in the
main experiments, Section~\ref{sec:method:alpha} selects
$(\alpha_{\mathrm{gain}},\alpha_{\mathrm{ugbw}},\alpha_{\mathrm{power}})
=(5,10,10)$ on four circuits and $(1,3,2)$ on 2-stage csmiller.
Section~\ref{sec:exp:alpha} reports the shared-default ablation.

\paragraph{Optimization parameters.}
Table~\ref{tab:hyperparams} lists the default ZO-MC-SGD settings.

\begin{paperappendixtable}
\centering
\caption{Default optimization hyperparameters.}
\label{tab:hyperparams}
\ifdefined\ARXIVVERSION
  \scriptsize
  \setlength{\tabcolsep}{2pt}
\else
  \footnotesize
  \setlength{\tabcolsep}{6pt}
\fi
\renewcommand{\arraystretch}{1.15}
\begin{tabular}{l l l}
\toprule
\rowcolor{tabBandYield}
\textbf{Symbol} & \textbf{Meaning} & \textbf{Default} \\
\midrule
$\epsilon$ & finite-difference step (normalized) & $5 \times 10^{-3}$ \\
\rowcolor{tabRowAlt}
$\eta$ & Adam step size & $0.05$ \\
$(\beta_1, \beta_2)$ & Adam moments & $(0.9, 0.999)$ \\
\rowcolor{tabRowAlt}
$K$ & ZO-MC-SGD inner $\boldsymbol{\xi}$ batch & $4$ \\
$N_{\mathrm{ws}}$ & warm-start size & $\max(1, \min(10, \lfloor B/16 \rfloor))$ \\
\rowcolor{tabRowAlt}
$n_{\mathrm{mc}}^{\text{ws}}$ & warm-start MC & $4$ \\
$n_{\mathrm{mc}}^{\mathrm{eval}}$ & final yield-evaluation MC & $80$ \\
\rowcolor{tabRowAlt}
$\Pi_{\mathcal{X}}$ & projection & clip onto $[0,1]^n$ \\
\bottomrule
\end{tabular}
\end{paperappendixtable}

\paragraph{RobustAnalog implementation.}
We implement RobustAnalog following the original
paper~\cite{shi2022robustanalog}: deep deterministic policy gradient
with multi-task gradient surgery and $k$-means corner pruning,
optimizing a specification-margin reward.

\paragraph{Coordinate conventions.}
All design-side quantities
($\epsilon$, $\eta$, $\Pi_{\mathcal{X}}$) act in the normalized
$\mathcal{X} = [0, 1]^n$ coordinates of Section~\ref{sec:method};
the per-benchmark $\sigma$-scale, $\alpha$ and starting design are
the ones released with the code.

\section{Additional Results}
\label{app:additional-results}

\paragraph{Optimizer overhead.}
Table~\ref{tab:overhead} reports optimizer-side runtime on the three CS
cascades at $B=1600$. ZO-MC-SGD, CMA-ES, PSO, and RobustAnalog add
little computation relative to SPICE, whereas BO and TuRBO incur
substantial surrogate-model and acquisition overhead as the design
dimension increases. At $n=30$, their optimizer times are $464$\,s and
$251$\,s, respectively, compared with approximately $350$\,s spent in
SPICE.

\begin{paperappendixtable}
\centering
\caption{Optimizer overhead on the CS cascades at $B=1600$. Median over
three seeds, excluding SPICE (${\approx}350$\,s for every method).}
\label{tab:overhead}
\footnotesize
\setlength{\tabcolsep}{5pt}
\renewcommand{\arraystretch}{1.15}
\begin{tabular}{l c c c c}
\toprule
\rowcolor{tabBandYield}
\textbf{Method} & $n{=}6$ & $n{=}18$ & $n{=}30$ & \textbf{\% total} ($n{=}30$) \\
\midrule
ZO-MC-SGD (proposed) & $40$\,ms & $43$\,ms & $48$\,ms & $0.01$ \\
\rowcolor{tabRowAlt}
BO & $121$\,s & $255$\,s & $464$\,s & $57.0$ \\
CMA-ES & $0.54$\,s & $0.55$\,s & $0.50$\,s & $0.1$ \\
\rowcolor{tabRowAlt}
PSO & $4$\,ms & $5$\,ms & $5$\,ms & $0.0$ \\
TuRBO & $62$\,s & $174$\,s & $251$\,s & $41.7$ \\
\rowcolor{tabRowAlt}
RA & $5.5$\,s & $5.9$\,s & $6.0$\,s & $1.7$ \\
\bottomrule
\end{tabular}
\end{paperappendixtable}

\paragraph{Fixed-budget yield.}
Table~\ref{tab:yield400} complements the minimum-budget comparison in
Table~\ref{tab:minbudget} by reporting the best mean yield attained
within $B=400$ simulations, together with the observed range over five
seeds.

\begin{paperwidetable}[t]
\centering
\caption{Best mean yield within a 400-simulation budget. Five-seed
mean with the observed range over the five seeds as a subscript;
RA = RobustAnalog.}
\label{tab:yield400}
\footnotesize
\ifdefined\ARXIVVERSION
  \setlength{\tabcolsep}{4pt}
\else
  \setlength{\tabcolsep}{6pt}
\fi
\renewcommand{\arraystretch}{1.25}
\begin{tabular}{l c c c c c c}
\toprule
\rowcolor{tabBandYield}
\textbf{Circuit ($d_\xi$)} & \textbf{ZO-MC-SGD} & \textbf{BO} &
\textbf{CMA-ES} & \textbf{PSO} & \textbf{TuRBO} & \textbf{RA} \\
\midrule
1-stage csamp (10) & \textbf{0.95}\,${}_{[0.95,\,0.95]}$ & \textbf{0.95}\,${}_{[0.95,\,0.95]}$ & 0.76\,${}_{[0.00,\,0.95]}$ & \textbf{0.95}\,${}_{[0.95,\,0.95]}$ & \textbf{0.95}\,${}_{[0.95,\,0.95]}$ & 0.52\,${}_{[0.24,\,0.95]}$ \\
\rowcolor{tabRowAlt}
3-stage csamp (26) & \textbf{1.00}\,${}_{[1.00,\,1.00]}$ & 0.83\,${}_{[0.20,\,1.00]}$ & 0.78\,${}_{[0.00,\,0.99]}$ & 0.94\,${}_{[0.86,\,1.00]}$ & 0.41\,${}_{[0.20,\,1.00]}$ & 0.52\,${}_{[0.20,\,0.99]}$ \\
5-stage csamp (42) & \textbf{1.00}\,${}_{[1.00,\,1.00]}$ & 0.60\,${}_{[0.34,\,1.00]}$ & 0.74\,${}_{[0.00,\,1.00]}$ & 0.87\,${}_{[0.34,\,1.00]}$ & 0.47\,${}_{[0.34,\,1.00]}$ & 0.34\,${}_{[0.34,\,0.34]}$ \\
\rowcolor{tabRowAlt}
2-stage csmiller (26) & \textbf{1.00}\,${}_{[1.00,\,1.00]}$ & \textbf{1.00}\,${}_{[1.00,\,1.00]}$ & \textbf{1.00}\,${}_{[1.00,\,1.00]}$ & \textbf{1.00}\,${}_{[1.00,\,1.00]}$ & \textbf{1.00}\,${}_{[1.00,\,1.00]}$ & 0.69\,${}_{[0.00,\,1.00]}$ \\
3-stage csmiller (38) & \textbf{1.00}\,${}_{[1.00,\,1.00]}$ & \textbf{1.00}\,${}_{[1.00,\,1.00]}$ & 0.98\,${}_{[0.93,\,1.00]}$ & \textbf{1.00}\,${}_{[1.00,\,1.00]}$ & 0.85\,${}_{[0.59,\,1.00]}$ & 0.93\,${}_{[0.66,\,1.00]}$ \\
\bottomrule
\end{tabular}
\end{paperwidetable}

\ifdefined\ARXIVVERSION
\section*{Acknowledgment}
The authors thank the National Institute of Standards and Technology (NIST) for supporting this work under Award \#70NANB24H084.
\else
\fi

\ifdefined\ARXIVVERSION
\bibliographystyle{IEEEtran}
\else
\bibliographystyle{ACM-Reference-Format}
\fi
\bibliography{refs}

\end{document}